\documentclass[twocolumn,trackchanges]{aastex701}

\begin{document}

\title{\large An H$\alpha$ Cloud in the HI Tail: Recent Star Formation in the Outskirts of NGC 4258 Revealed by Nanshan 1-m Telescope}

\author[0000-0003-0202-0534]{Cheng Cheng}
\affiliation{Chinese Academy of Sciences South America Center for Astronomy, National Astronomical Observatories, CAS, Beijing 100101, People’s Republic of China, chengcheng@nao.cas.cn}
\affiliation{CAS Key Laboratory of Optical Astronomy, National Astronomical Observatories, Chinese Academy of Sciences, Beijing 100101, People’s Republic of China}
\email[show]{chengcheng@nao.cas.cn}  

\author[0000-0001-6511-8745]{Jia-Sheng Huang}
\affiliation{Chinese Academy of Sciences South America Center for Astronomy, National Astronomical Observatories, CAS, Beijing 100101, People’s Republic of China, chengcheng@nao.cas.cn}
\affiliation{CAS Key Laboratory of Optical Astronomy, National Astronomical Observatories, Chinese Academy of Sciences, Beijing 100101, People’s Republic of China}
\affiliation{Harvard-Smithsonian Center for Astrophysics, 60 Garden Street, Cambridge, MA 02138, USA}
\email{jhuang@nao.cas.cn}

\author[0000-0002-5674-4223	]{Peng Wei}
\affiliation{Xinjiang Astronomical Observatory, Chinese Academy of Sciences, Urumqi, Xinjiang 830011, P. R. China, weipeng@xao.ac.cn, aliyi@xao.ac.cn}
\affiliation{University of Chinese Academy of Sciences, No.19(A) Yuquan Road, Shijingshan District, Beijing 100049, P. R. China}
\email[show]{weipeng@xao.ac.cn}

\author[0000-0003-1845-4900]{Ali Esamdin}
\affiliation{Xinjiang Astronomical Observatory, Chinese Academy of Sciences, Urumqi, Xinjiang 830011, P. R. China, weipeng@xao.ac.cn, aliyi@xao.ac.cn}
\affiliation{University of Chinese Academy of Sciences, No.19(A) Yuquan Road, Shijingshan District, Beijing 100049, P. R. China}
\email[show]{aliyi@xao.ac.cn}

\author[0000-0002-4788-8188]{Guojie Feng}
\affiliation{Xinjiang Astronomical Observatory, Chinese Academy of Sciences, Urumqi, Xinjiang 830011, P. R. China, weipeng@xao.ac.cn, aliyi@xao.ac.cn}
\affiliation{University of Chinese Academy of Sciences, No.19(A) Yuquan Road, Shijingshan District, Beijing 100049, P. R. China}
\email{fengguojie@xao.ac.cn}

\author[0000-0002-2419-6875]{Zhi-Xiang Zhang}
\affiliation{College of Physics and Information Engineering, Quanzhou Normal University, Quanzhou, Fujian 362000, People’s Republic of China, zzx@qztc.edu.cn}
\email{zzx@qztc.edu.cn}

\author[0000-0001-7592-7714]{Haojing Yan} 
\affiliation{Department of Physics and Astronomy, University of Missouri, Columbia, MO 65211}
\email{yanhaojing@gmail.com}

\author[0000-0003-4546-8216]{Wei Du}
\affiliation{CAS Key Laboratory of Optical Astronomy, National Astronomical Observatories, Chinese Academy of Sciences, Beijing 100101, People’s Republic of China}
\affiliation{National Astronomical Observatories, Chinese Academy of Sciences, Beijing 100101, China}
\email{wdu@nao.cas.cn}

\author[0000-0003-3948-9192]{Pei Zuo}
\affiliation{National Astronomical Observatories, Chinese Academy of Sciences, Beijing 100101, China}
\email{peizuo@nao.cas.cn}

\author[0000-0001-7634-1547]{Zi-Jian Li}
\affiliation{Chinese Academy of Sciences South America Center for Astronomy (CASSACA), \\
National Astronomical Observatories of China (NAOC),\\
CAS, 20A Datun Road, Beijing 100012, China}
\affiliation{School of Astronomy and Space Sciences, University of Chinese Academy of Sciences, Beijing 100049, China}
\email{zjli@nao.cas.cn}

\author[0000-0002-6642-7483]{Gustavo Orellana}
\affiliation{Fundaci\'on Chilena de Astronom\'ia, c\'odigo postal 7500011, Santiago, Chile}
\email{gustavo.orellana.gonzalez@gmail.com}

\author{Letian Wang}
\affiliation{Xinjiang Astronomical Observatory, Chinese Academy of Sciences, Urumqi, Xinjiang 830011, P. R. China, weipeng@xao.ac.cn, aliyi@xao.ac.cn}
\affiliation{University of Chinese Academy of Sciences, No.19(A) Yuquan Road, Shijingshan District, Beijing 100049, P. R. China}
\email{wangletian@xao.ac.cn}

\author{Yong Wang}
\affiliation{Xinjiang Astronomical Observatory, Chinese Academy of Sciences, Urumqi, Xinjiang 830011, P. R. China, weipeng@xao.ac.cn, aliyi@xao.ac.cn}
\affiliation{University of Chinese Academy of Sciences, No.19(A) Yuquan Road, Shijingshan District, Beijing 100049, P. R. China}
\email{wangyong@xao.ac.cn}

\author[0009-0003-9229-9942]{Abdusamatjan Iskandar}
\affiliation{Xinjiang Astronomical Observatory, Chinese Academy of Sciences, Urumqi, Xinjiang 830011, P. R. China, weipeng@xao.ac.cn, aliyi@xao.ac.cn}
\affiliation{University of Chinese Academy of Sciences, No.19(A) Yuquan Road, Shijingshan District, Beijing 100049, P. R. China}
\email{abudu@xao.ac.cn}

\author{Shahidin Yaqup}
\affiliation{Xinjiang Astronomical Observatory, Chinese Academy of Sciences, Urumqi, Xinjiang 830011, P. R. China, weipeng@xao.ac.cn, aliyi@xao.ac.cn}
\affiliation{University of Chinese Academy of Sciences, No.19(A) Yuquan Road, Shijingshan District, Beijing 100049, P. R. China}
\email{xiayiding@xao.ac.cn}

\begin{abstract}

We present first-light deep H$\alpha$ imaging taken with the Nanshan 1-meter wide-field telescope on the local galaxy NGC 4258, alongside archival data from Hubble Space telescope (HST), Westerbork Synthesis Radio Telescope, and The Dark Energy Camera Legacy Survey. The H$\alpha$ image shows ongoing star formation not only inside the galaxy but also in an HI cloud in the eastern HI tail, which is roughly 16 kpc away from the main galaxy. The HST images reveal several ultra-blue compact objects ($\rm F555W - F814W <-0.5 mag,\, FWHM\sim 0.2''$) in the H$\alpha$ bright region, aligned with the HI tail, suggesting the presence of young open cluster candidate in the HI tail. Our results suggest that wide field H$\alpha$ imaging is a valuable tool for investigating recent star formation in the extended regions of NGC 4258. Furthermore, the star formation in diffuse HI tails could highlight an potential aspect of galaxy stellar halo formation, warranting further investigation of the impact of star formation in halos on galaxy evolution.

\end{abstract}



\section{Introduction} \label{sec:intro}

Deep H$\alpha$ images of galaxies are widely used to trace recent high-mass star formation events occurring within the past 10 Myr \citep{2012ARA&A..50..531K, 2012AJ....144....3L}. The H$\alpha$ emission is one of the most accurate and widely used indicators of star formation rate (SFR), as its wavelength falls within the optical range and can be efficiently observed with optical spectroscopy \citep{1998ARA&A..36..189K}. To identify individual HII regions across a galaxy, H$\alpha$ narrow-band imaging has emerged as the primary method for tracing recent high-mass star formation, leveraging its wide-field observational capabilities and effective separation of emission-line signatures. On the other hand, although spectroscopic methods can also provide measurements of the total star formation activity in galaxies (e.g., \citealt{2000ApJS..126..331J}), H$\alpha$ narrow-band imaging offers a more efficient and observationally straightforward approach for estimating galaxy-wide SFRs (e.g., 
\citealt{1973ApJ...185..869R};
\citealt{1992A&A...257..389L};
\citealt{1998PASA...15....9R};
WHAM, \citealt{1998PASA...15...14R};
\citealt{1998PASA...15..147D};
\citealt{2001PASP..113.1326G};
\citealt{2002A&A...387..821G};
\citealt{2003PASP..115..928K}; 
GOLDMine, \citealt{2003A&A...400..451G};
\citealt{2004A&A...414...23J};
SINGG, \citealt{2006ApJS..165..307M};
11HUGS, \citealt{2008ApJS..178..247K};
MOSAIC, \citealt{2012ApJS..199...36S};
\citealt{2018ApJS..235...18L}; 
\citealt{2020AJ....160..242S};
\citealt{2022ApJ...927..136L};
KMTNet Nearby Galaxy Survey, \citealt{2022PASP..134i4104B};
\citealt{2024MNRAS.530.4988L};
\citealt{2024arXiv241001884D};
The PHANGS-HST-H$\alpha$ Survey, \citealt{2025AJ....169..150C}).

In addition to successfully estimating the star formation rates within galaxies, H$\alpha$ emission has been observed extending up to $\sim$ 10 kpc beyond the main optical disk in a few local galaxies such as M51 \citep{2018ApJ...858L..16W}, M81 Group \citep{2022ApJ...927..136L, 2024arXiv241106258L, 2024arXiv241106255L}, M31 \citep{2018ApJ...853...50F}, NGC 3344 \citep{2021ApJ...923..199P}, NGC 7331 \citep{2007ApJS..173..572T}, Sculptor group galaxies NGC 253 \citep{2011MNRAS.411...71H}, NGC 247 \& 300 \citep{2011MNRAS.416..509H}, and NGC 1068 \citep{2024ApJ...974..161M}, which are likely to be a hint of black hole activity, just like Hanny's Voorwerp \citep{2009MNRAS.399..129L, 2022MNRAS.510.4608K}. Other common extended H$\alpha$ emission is in jelly fish galaxies in galaxy clusters such as Virgo \citep{2016A&A...587A..68B, 2008ApJ...687L..69K}, Coma \citep{2010AJ....140.1814Y, 2022NatAs...6..270S}, A1367 \citep{2021MNRAS.505.4702G}, which are the consequence of ram pressure or tidal disruption. Moreover, diffuse H$\alpha$ emission is also identified from spectroscopic observations \citep{2019MNRAS.485..428B, 2021MNRAS.508.3943J, 2024ApJ...972L..16Y, 2024arXiv241005392Z}, indicating a different ionization process. 

\begin{figure}
    \centering
    \includegraphics[width=0.98\linewidth]{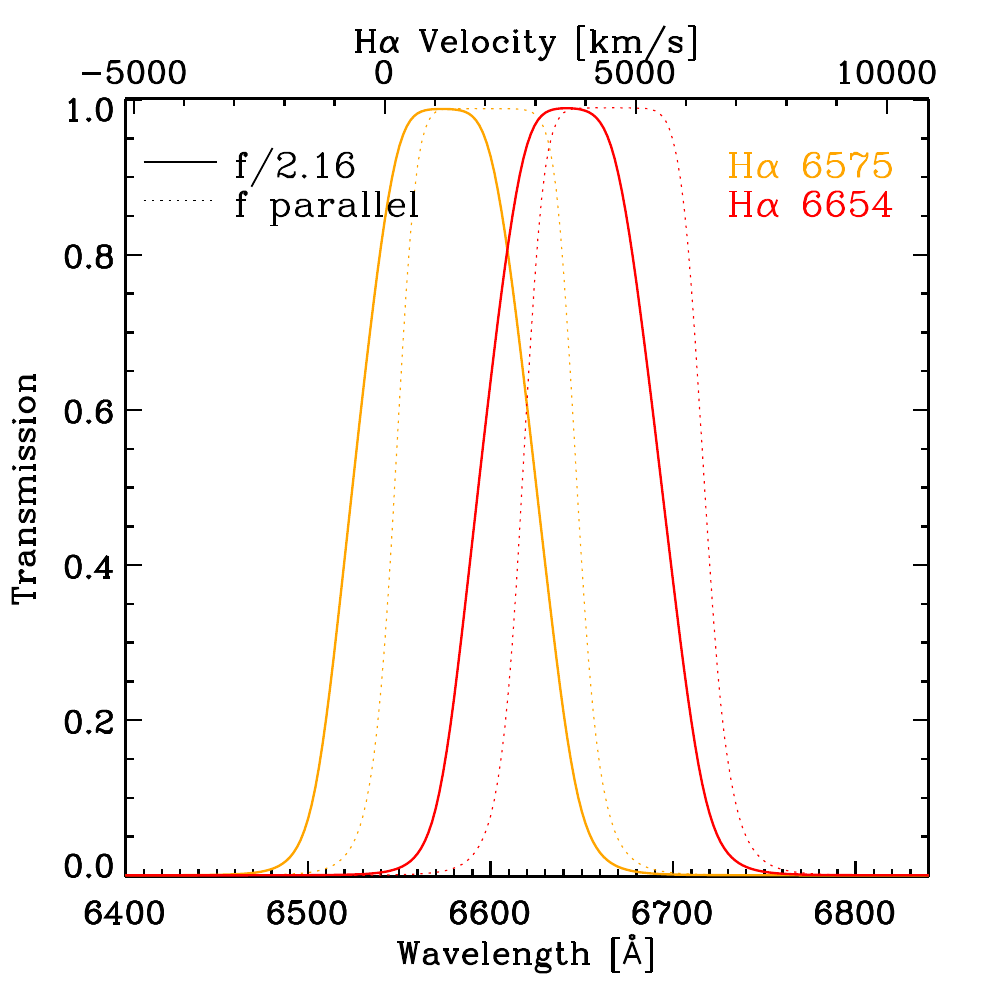}
    \caption{Transmission curve of the narrow band filters. The H$\alpha$ observation of NGC 4258 is taken by the H$\alpha$ filter with the central wavelength of 6575\AA. 
    The dotted curves are the transmission curves of the filters under parallel beam, and the solid lines are under the f/2.16 focus beam estimated from the parallel curve. 
    The data for all curves are provided by the ASAHI Corporation. The filters are installed in the converging beam, so the central wavelengths shift toward the blue side.
    }
    \label{filter}
\end{figure}

Most bright local galaxies have been imaged through H$\alpha$ or [OIII] narrowband filters of various prescriptions to address a host of specific scientific questions. Consequently, these narrowband images reach a range of depths, provide often limited areal coverage within galaxies, and render uniform study of galaxy properties difficult. In order to trace star formation via H$\alpha$ emission and do so over the full extent of nearby galaxies, two H$\alpha$ narrowband filters were procured from ASAHI Corporation (Figure \ref{filter}) and installed in the prime focus camera of the Nanshan 1-meter Wide Field Telescope \citep{2020RAA....20..211B}. The primary science goals are to identify H$\alpha$ bright regions in the outer regions of local galaxies to understand the star formation in galactic halos. 
The sample we selected are the local galaxies with archival multi-wavelength data, especially with high resolution HI data from e.g., The Westerbork Hydrogen Accretion in LOcal GAlaxieS \citep[HALOGAS, ][]{2011A&A...526A.118H}, The HI Nearby Galaxy Survey \citep[THINGS,][]{2008AJ....136.2563W} and MeerKAT H I Observations of Nearby Galactic Objects: Observing Southern Emitters \citep[MHONGOOSE,][]{2024A&A...688A.109D} projects.

We choose NGC 4258 as the first-light target for the newly installed 6575 \AA\ H$\alpha$ filter. NGC 4258 serves as an ``anchor" galaxy for distance measurements \citep[7.6 Mpc,][]{2019ApJ...886L..27R, 2024ApJ...977..120R}, which is within the wavelength coverage of our H$\alpha$ filter wavelength coverage (Figure \ref{filter}). The optical diameter of NGC4258 is about 22 arcmin \citep{https://doi.org/10.26132/ned1}\footnote{\url{https://ned.ipac.caltech.edu/byname?objname=ngc4258}}, which is suitable for the 1 degree Field of View of Nanshan 1m telescope. NGC 4258 has 
an accurate super massive black hole mass measurement from its maser \citep{1994ApJ...437L..35H,1995Natur.373..127M,1995ApJ...440..619G,1999Natur.400..539H}. The super massive black hole activity also leads to a strong jet, shock and outflow, as revealed by X-ray and ionized gas \citep{1993A&A...268..419C, 2000ApJ...536..675C, 2001ApJ...560..689W, 2010MNRAS.406..181J, 2018ApJ...869...61A, 2022A&A...663A..87M, 2023MNRAS.526..483Z}. The HI observation from Westerbork Synthesis Radio Telescope (WSRT) shows two HI tails in the east and north of the main body of the galaxy (Figure \ref{stamp}). Recently, FAST observations revealed a 100 kpc long HI stream, indicating a complex interaction between NGC4258 and the neighboring galaxies \citep{2021ApJ...922L..21Z, 2024MNRAS.534.3688L}. The well-studied properties, multi-wavelength archival data, declination, redshift, and know high H$\alpha$ brightness all make NGC 4258 an ideal first-light target.

Previous H$\alpha$ observations of NGC 4258 \citep[e.g., ][]{1970A&A.....9..181D, 1993A&A...268..419C,1986ApJ...311L...7F, 1989ApJ...345..707M} were focusing on the central region of the galaxy to investigate the jet, shock, anomalous arms, etc. Thanks to the wide field coverage of Nanshan 1m telescope primary focus camera, one advantage of our H$\alpha$ observations of NGC 4258 is that they cover not only the main body of the galaxy but also its outskirt region in single pointings.

In this study, we report the discovery of an H$\alpha$-emitting region identified in our H$\alpha$ image of NGC 4258. The H$\alpha$ cloud coincides with an HI tail that is $\sim$16~kpc away from the galaxy's center in project distance. 
We also identify several young, open cluster candidate in the archival
HST images that appear to be co-located with the H$\alpha$-bright region,
implying a recent star formation event in the HI tail or even in the
circumgalactic medium of NGC 4258. Throughout the paper, we adopt the Chabrier IMF \citep{Chabrier2003} and a standard Lambda cold dark matter cosmology ($\Lambda$CDM) with $\rm \Omega_m=0.3$, $\rm \Omega_\Lambda=0.7$, and H$\rm _0=70\, km\,s^{-1}\,Mpc^{-1}$, 

\begin{figure*}
    \centering
    \includegraphics[width=0.98\linewidth]{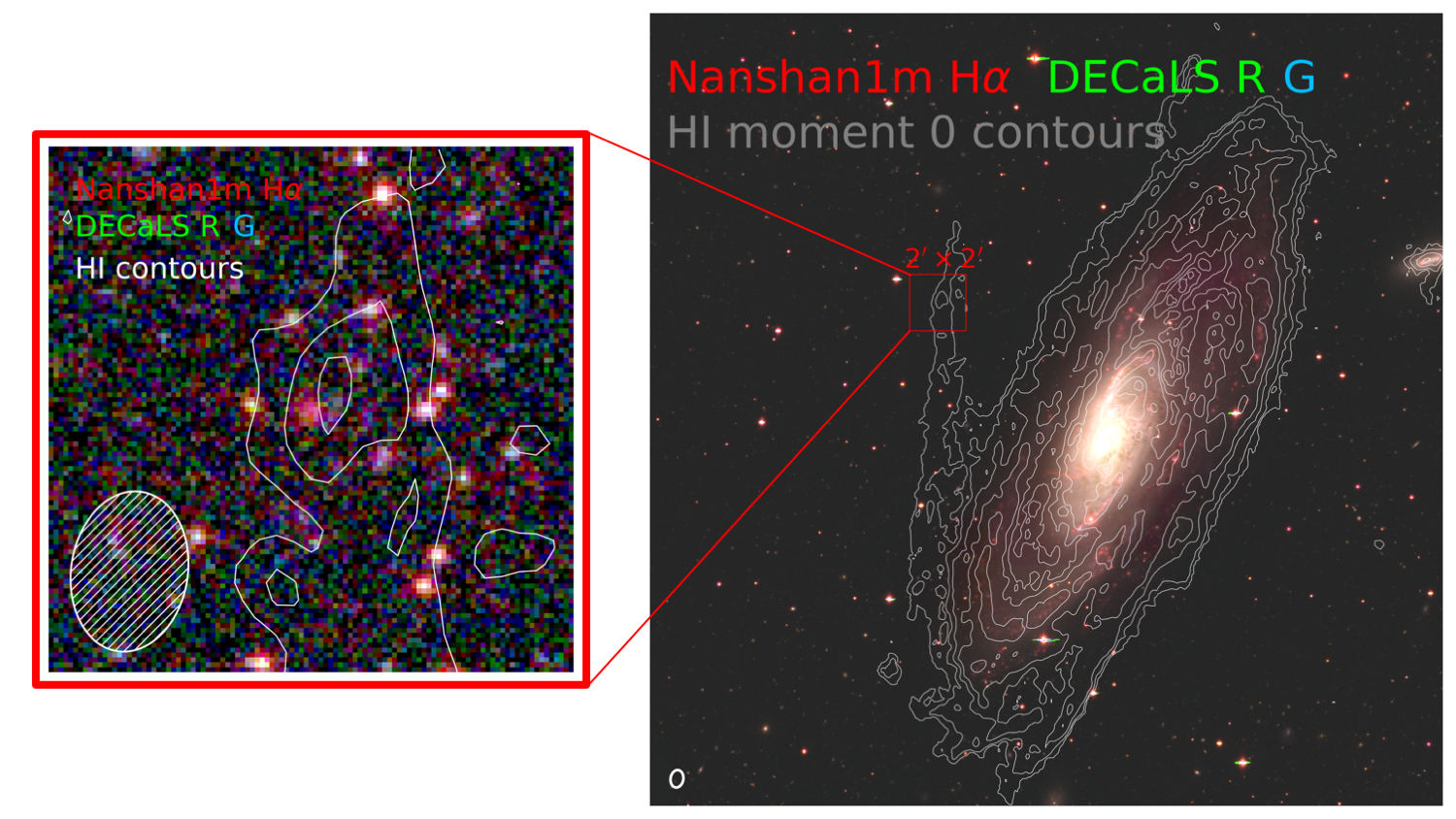}
    \caption{{\bf Right panel}:  
    False color composite of the Nanshan 1m H$\alpha$ 
    narrow band image (red), DECaLS R (green) and G (blue) band images ($28'\times 28'$; image top is north and left is east). We make use of the {\sc ASINH} stretch to highlight the H$\alpha$ emission in the galaxy center and galactic disk, while the H$\alpha$ emission at the extended region is suppressed. We overplot the HI flux contours from the HI moment 0 image produced by HALOGAS project. The contours start from 0.01 Jy/Beam km/s, and the HI data beam is shown in he lower left corner with white ellipse. The detected H$\alpha$ emitter locate in the HI tail and highlighted by a red box  ($2'\times2'$). A zoomed-in version of which is shown in the {\bf left panel}: color image with H$\alpha$ (red), R (green) and G (blue), centered at RA = 12:19:33.3821, Dec = +47:23:05.509 (J2000). To highlight the HI map in the tail, 
    we produce the HI moment 0 map by integrating the data cube along the velocity axis
    within 405 to 459 km$\,\rm s^{-1}$, and shown in white contours with HI column density $[5, 10, 15, 20, ...] \times 6.4\times 10^{20} \rm cm^{-2}$. The complete moment 0 map for the HI tail is shown in Figure \ref{HIm0}.The white hatched ellipse at the lower left corner represents the synthesized beam. 
    }
    \label{bigpic}
\end{figure*}

\begin{figure*}
    \centering
    \includegraphics[width=0.95\linewidth]{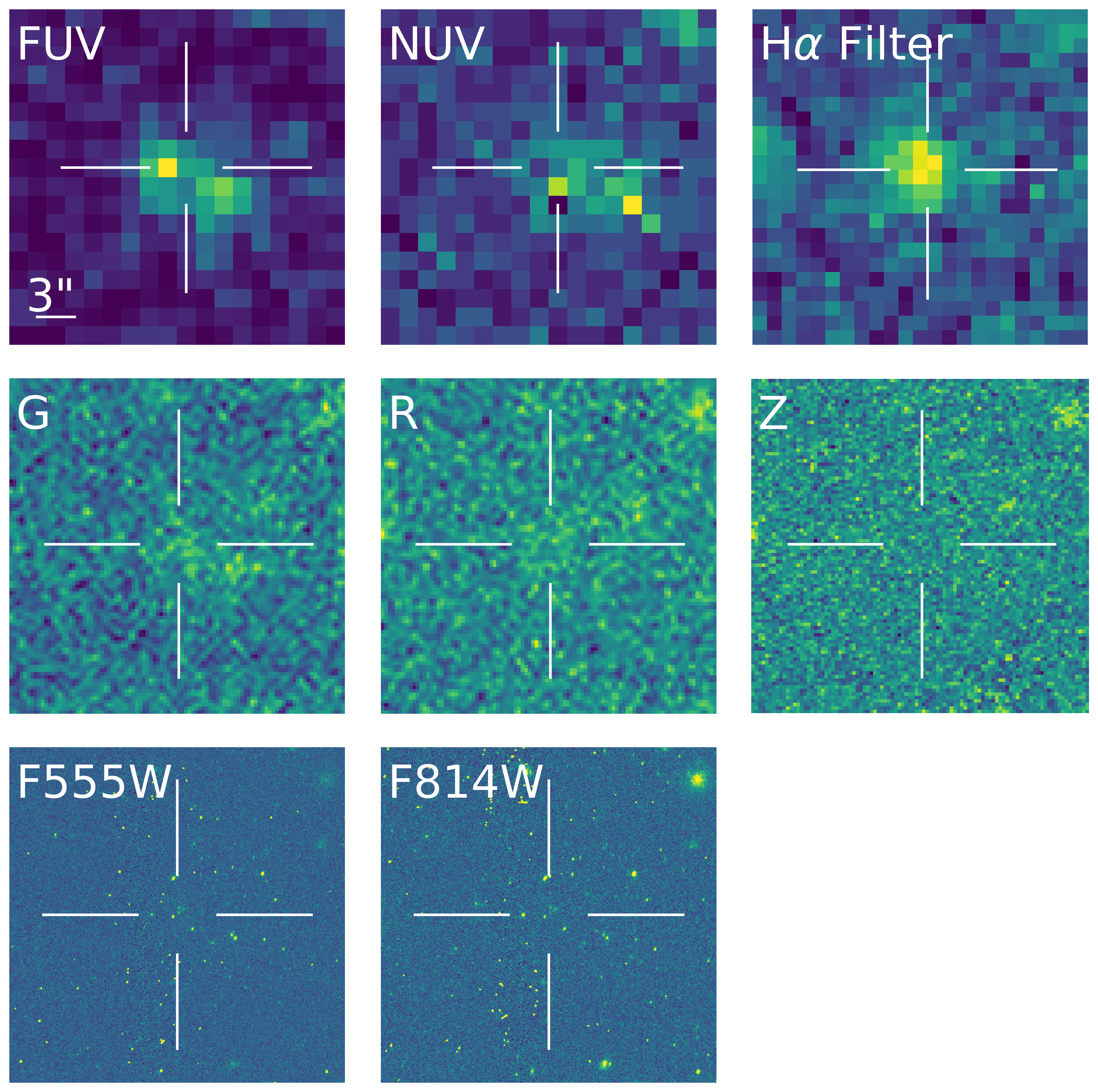}
    \caption{Multi-wavelength stamp images in GALEX/FUV, GALEX/NUV, H$\alpha$ narrow band filter, DECaLS/G/R/Z, HST/F555W, HST/F814W bands, centered on the coordinate of the H$\alpha$ emitter. The size of each image is $50''\times50''$. The depth of the DECaLS images are about 24 AB mag, and the H$\alpha$ emitter in G band is only marginally detected. The images are centered at RA = 12:19:33.3821, Dec = +47:23:05.509 (J2000).
    }
    \label{stamp}
\end{figure*}

\section{Observation, archival data and the H$\alpha$ emitter}

\subsection{Transmission curve of the H$\alpha$ narrow band filters in Nanshan 1m telescope}

We show the transmission curve of our narrow band filters installed in the Nanshan 1m telescope in Figure \ref{filter}. The NGC 4258 H$\alpha$ image presented in this work is taken by the H$\alpha$ filter with the central wavelength 6575\AA\ (orange curve in Figure \ref{filter}). The peak transmission is 98.9\%. The filters are installed in the converging beam,  so the central wavelength is shifted compared to that in a collimated (or parallel) beam \citep[e.g., ][]{2019PASP..131g4502Z}. 
The central wavelength of the two filters are 6575 and 6654\AA\, and the full width at half maximum (FWHM) is 100\AA. The two filters can cover the targets at $0<z<0.015$.
To avoid the shifts of the transmission central wavelength across the full 1.4-degree field of view, we only use the central $40' \times 40'$ region of the full CCD.

\subsection{H$\alpha$ observation and reduction}

We imaged NGC 4258 during 2022 April with the prime focus camera of the
Nanshan 1-meter Wide Field Telescope \citep{2020RAA....20..211B} using an H$\alpha$
narrowband filter with a central wavelength of 6575 \AA\ and a width of
100\AA\ (Figure \ref{filter}). The primary focus camera yields a field of view of 1.4 deg (f/2.16) and the pixel scale is about 1.4 arcsec. The typical seeing at Nanshan Observatory is about 2. to 3 arcseconds. This site is primarily designed for radio observations, so its optical seeing is not as optimal as other observatories. To reduce the distortion at the edge of the image and save the readout overhead, we only readout the central $40' \times 40'$. 
We readout each image with the slowest speed, and the readout noise is 3.1 e$^{-1}$.
The telescope does not have a guide star system. To limit image smearing, we set the exposure time as 3 min each. We took bias frames for each night, and twilight flat frames when the weather is good. The dither pattern introduces a random 3'' offset between consecutive frames, which helps to remove the bad pixels robustly.

We calibrated all narrowband images using standard procedures with the nightly bias frames and flat exposures taken closest in time. We removed cosmic ray induced signal using
\textsc{la cosmic} \citep{2001PASP..113.1420V}, and calibrate the astrometry to GAIA DR2 \citep{2018A&A...616A...1G} with \textsc{astrometry.net} \citep{2010AJ....139.1782L}. The wide field of view enable a good control of background, which is removed by \textsc{Noisechisel} \citep{2015ApJS..220....1A}. We combine a pipeline to link all the basic steps, and to produce the images with bias, flat, astrometry calibrated, and sky background subtracted \footnote{\url{https://github.com/chengchengcode/nanshan1m-pipeline}}. Then we check images by eye to remove any problem images including poor guiding, too strong background caused by the Moon or thin cloud, strange readout pattern, or other issues.

We mosaic all the images together by \textsc{swarp} \citep{2002ASPC..281..228B}. 
The mosaic image has a pixel size of $1.133''$/pix., and a FWHM of $3.44''$.
We calibrate the image flux with DECaLS R band image \citep{2019AJ....157..168D}, which has the central wavelength of 6417 A. The total exposure time on NGC 4258 is 15.8 hours, resulting in a 3 $\sigma$ depth of 20.9 AB magnitude in a $2.5''$ diameter aperture.

\begin{figure*}
    \centering
    \includegraphics[width=0.98\linewidth]{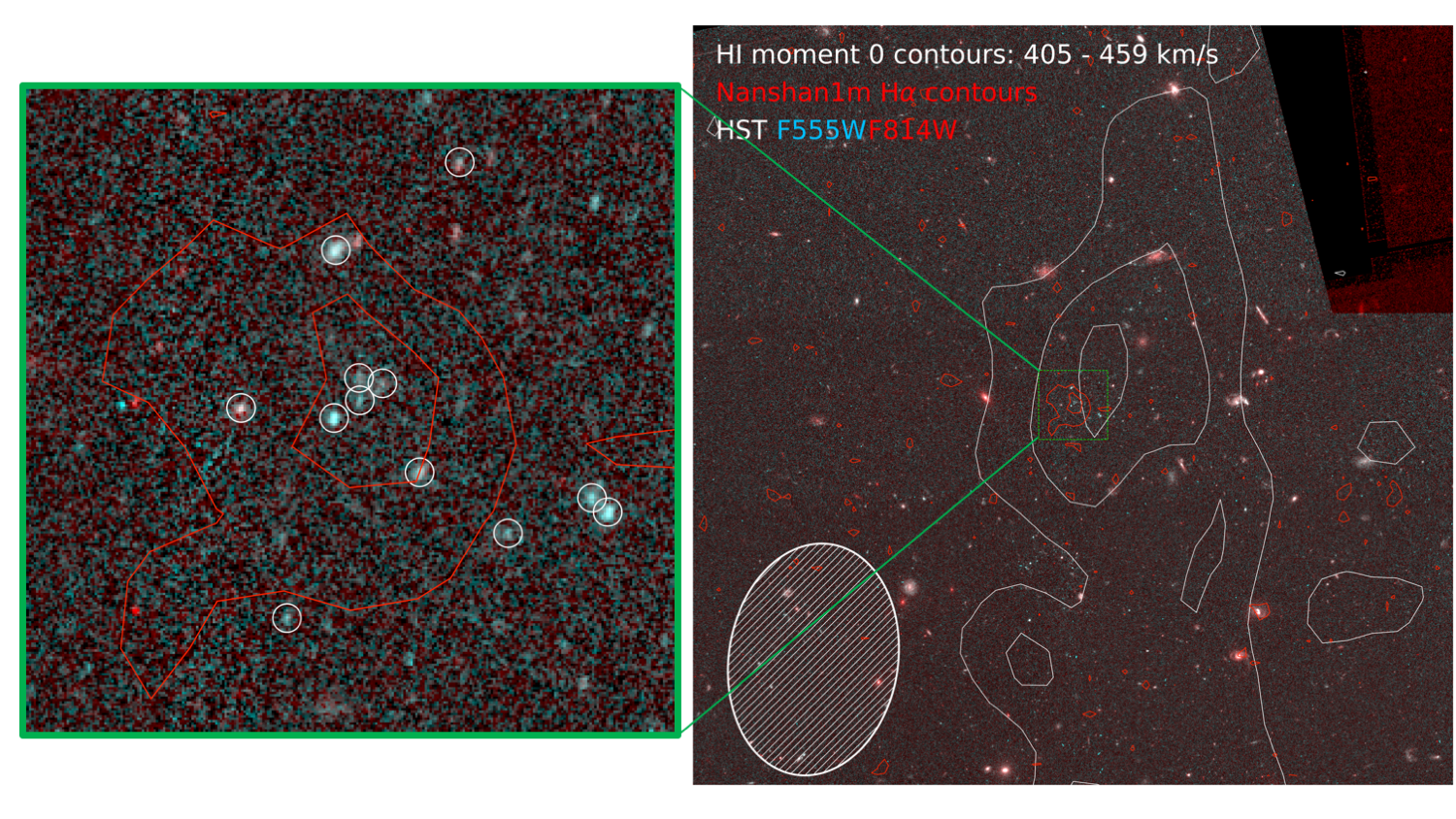}
    \includegraphics[width=0.95\linewidth]{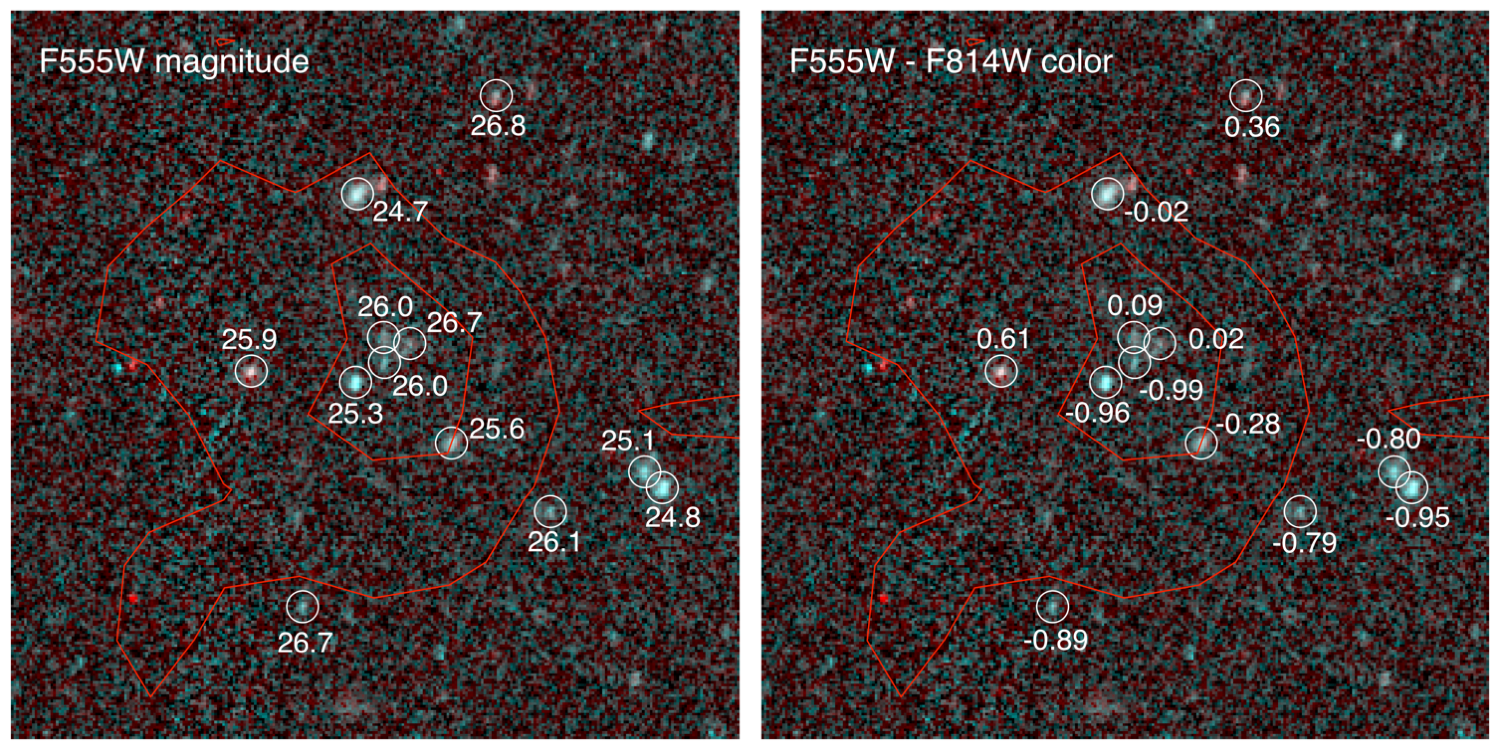}
    \caption{{\bf Top right panel}: HST color image with F555W (blue color), F814W (red color) bands, overploted with the HI contour in white color and H$\alpha$ contours in red.    
    The image size is $2'\times2'$. The HI moment 0 map is derived from 405 to 459 km/s, which is optimized by the HI emission in the HI knots in the HI tail. The green box is the central region ($11''\times11''$) of the H$\alpha$ emitter, and zoom in at the {\bf top left panel.}
    {Lower left panel: } Same image as top left panel, with F555W magnitude denoted. {Lower panels: } Same image as the top left panel, with the F555W AB magnitude denoted in lower left panel and the F555W - F814W color denoted in lower right panel. The H$\alpha$ emitter region includes targets with F555W about 25 - 26 AB mag, and F555W - F814W $ \lesssim 0$, indicating a hot massive stellar population. The crowded blue targets in the H$\alpha$ region implies the recent star formation activity.
    }
    \label{HST}
\end{figure*}

\subsection{Archival data used in this work}

NGC 4258 has been observed by various survey projects across multiple wavelengths. We make use of the DECaLS images \citep{2019AJ....157..168D} in G, R, Z bands to compare with the H$\alpha$ image (Figure \ref{bigpic}), The depth of the DECaLS images are about 24, 23 an 22 AB mag in G, R and Z band respectively \citep[more details of the survey design and depth of the DECaLS can be found in ][]{2019AJ....157..168D}. Other datasets we make use of in this study are summarized below.

The WSRT data for NGC 4258 is part of the HALOGAS project \citep{2011A&A...526A.118H}, which observed a sample of local galaxies with an integration time of about 120 hours each. The observations reach a neutral hydrogen column density of $4 \times 10^{18}\, \mathrm{cm^{-2}}$ at a velocity resolution of 4.2 km$\,\rm s^{-1}$. The calibrated HI cube and HI moment 0 map is publicly available on Zenodo \citep{heald_2020_3715549}\footnote{\url{https://zenodo.org/record/3715549\#.X2XQ4ZMzZ24}}. The synthesized beam size of the cube is $20.9'' \times 14.7''$, with a position angle of $-8.78^\circ$, and the velocity bin width
is 4.2 km$\,\rm s^{-1}$. To highlight the HI emission in the HI tail, we build the HI moment 0 map within 405 to 459 km$\,\rm s^{-1}$, which is the velocity range in the HI tail, and shown the HI moment 0 map in white contours in the left panel of Figure \ref{bigpic}, the upper right panel in Figure \ref{HST} and Figure \ref{HIm0}.

{\it Hubble} Space Telescope (HST) observations that cover the H$\alpha$-bright region were conducted between 2003 and 2005 using the ACS/F555W and ACS/F814W filters. These images were used to calibrate the galaxy's distance via the Tip of the Red Giant Branch method \citep{2021ApJ...906..125J}. We downloaded all ACS image mosaics covering the emitter from the MAST archive \footnote{\url{https://mast.stsci.edu/portal/Mashup/Clients/Mast/Portal.html}}. These images, reduced and combined into mosaics using the default pipeline, have a pixel scale of $0.04''$ and a 5$\sigma$ depth of approximately 26 AB magnitudes, depending on the observation coverage. The adopted zero point of the ACS images is $\rm ZP_{\rm AB} = -2.5 \log(PHOTFLAM) - 5\log(PHOTPLAM) - 2.408 $ mag \footnote{\url{https://www.stsci.edu/hst/instrumentation/acs/data-analysis/zeropoints}}. Cosmic rays in the HST images were removed using \textsc{L.A.Cosmic} \citep{2001PASP..113.1420V}.

We also utilized Galaxy Evolution Explorer \citep[GALEX,][]{2005ApJ...619L...1M} FUV and NUV images of NGC 4258. The GALEX images of the H$\alpha$ emitter are downloaded from the GALEX archive website \footnote{\url{https://galex.stsci.edu/gr6/}} with the tile name $\rm PS\_NGC4258\_MOS23$. The zeropoint is 18.82 for FUV image and 20.08 for NUV image.

All the HST and GALEX data used in this paper can be found in MAST: \dataset[https://doi.org/10.17909/jhga-gy40]{https://doi.org/10.17909/jhga-gy40}.

\subsection{The H$\alpha$ emitter in the eastern HI tail of NGC 4258}

A composite false-color image from H$\alpha$, R, G filters reveals a distinct red emitter located approximately 16 kpc projected from the center of NGC 4258 (Figure \ref{bigpic}). This feature represents the primary discovery of this study. Comparing with the HI moment 0 map from WSRT, the H$\alpha$ emitter aligns well with the HI tail, suggesting that it is likely a star-forming region within the HI knots in the tail. At the current depth of our image, this is the only new H$\alpha$ emitter found.

To confirm the redshift of the H$\alpha$ emitter, we present zoomed-in stamp images in GALEX/FUV, GALEX/NUV, Nanshan 1m/H$\alpha$, DECaLS/G/R/Z, HST/F555W, and HST/F814W bands (Figure \ref{stamp}). This emitter is faint at the DECaLS G-band images, with no clear detections in the R and Z bands. This suggests that the stellar component is fainter than roughly 24 AB mag at optical wavelengths, which results in the red apparent color of the source in Figure \ref{bigpic}. The GALEX images show clear detections in the UV bands (Figure \ref{stamp}). Consequently, this emitter is unlikely to be a Ly$\alpha$ emitter at redshift 4.4, and more likely resides at low redshifts. The overlap of the H$\alpha$, UV, and HI contours strongly supports the idea that a H$\alpha$ emitter originates from the star forming region in HI tail. 

Although the optical continuum is not clearly detected in the DECaLS images, high-resolution HST images reveal several blue sources (Figure \ref{HST}) within the H$\alpha$ emission region (Using the DECaLS R-band image as the continuum, we subtracted the PSF-matched R-band image from the H$\alpha$ filter image. The resulting H$\alpha$ emission contours are shown in Figure \ref{HST}). These sources are identified as the optical counterparts of the H$\alpha$ emitter. Their F555W - F814W colors are approximately $-0.8$ mag, indicative of extremely blue O or B-type stars with $T_{\rm eff} \sim 20000$ K (Figure \ref{Teff}). The F555W magnitude is about 25 AB mag, making these sources undetectable in DECaLS images. 
If these stars resided within our own Galaxy at a distance of 10 kpc
then 25.5 AB mag would correspond to $\sim 0.4\, L_\odot$\footnote{We consider a star with an observed magnitude in F555W band as 25.5 AB at a distance of 10 kpc, and correct to the total luminosity using the TLUSTY stellar model \citep{zhang_2025_15109587, 2007ApJS..169...83L} with $T_{\rm eff} = 20{,}000$ K.}, inconsistent with
the luminosity of an O or B star. Meanwhile, the spatial clustering of the sources within the H$\alpha$ emitting region and H I knot strongly argues against Galactic white dwarfs as a possibility.

If the H$\alpha$ emitter is associated with NGC 4258, the total luminosity of the blue sources seen in the HST image is approximately $2.2 \times 10^5\,L_\odot$, assuming an F555W-band magnitude of 25.5 AB at a distance of 7.6 Mpc. This corresponds to a stellar mass of $\sim$ 32\,$M_\odot$ for a main-sequence star \citep{2015AJ....149..131E}, consistent with the presence of massive O- or B-type stars. And the discrete blue targets are likely to be a young open cluster formed in a knot of the HI tail.

Meanwhile, a supernova remnant is also a possible explanation for the observed H$\alpha$ emission \citep{1979ApJS...39....1R, 2015SerAJ.191...67V}. In this case, the H$\alpha$ emission is primarily shock-induced and unrelated to H II regions or recent star formation. Consequently, the star formation rate estimated from H$\alpha$ luminosity would be overestimated \citep{2011ApJ...741..124H, 2012ARA&A..50..531K}. 
The contamination from supernova remnants in estimating the star formation rate in galaxies is generally small \citep[e.g., Appendix B in ][]{2004MNRAS.351.1151B}, while its impact in this HI clouds remains uncertain.
Therefore, high resolution H$\alpha$ image, multi-wavelength spectral energy distribution (SED) from U to I band based on high resolution images, and spectroscopic observations are still needed to decipher the origin of the H$\alpha$ emission.

Another possible interpretation is that the emitter is an [OIII] emitter such as the extremely metal-poor galaxies \citep[EMPGs, ][]{2020ApJ...898..142K} at redshift 0.31. However, the discrete morphology observed in the HST image is rarely seen at this redshift. And the 25 AB mag in F555W at $z = 0.31$ would correspond to a total stellar mass $\sim 10^9 M_\odot$ \citep[based on the mass-light ratio given by][]{2003ApJS..149..289B}, which is about 2 to 3 orders of magnitude more massive than a typical EMPG. Based on these luminosity estimations and the overlap with the HI map, we conclude that the H$\alpha$ emitter is likely to be an HII region ionized by a young, open cluster within the HI tail of NGC 4258.

\subsection{Physical Properties of the emitter}

We performed photometry of the H$\alpha$ emitter using SExtractor \citep{1996A&AS..117..393B} and obtained an AUTO magnitude of 20.68 $\pm$ 0.07 mag. Since the H$\alpha$ emitter is not clearly detected in DECaLS images, we adopted the 5$\sigma$ limiting magnitudes in a 20" aperture: 22.4 AB mag in the R band and 23.1 AB mag in the G band. The line flux of the H$\alpha$ emitter was estimated using the equation:
\begin{equation}
    F_{\rm Line} = \Delta NB \frac{f_{\rm NB} - f_{\rm R}}{1-\Delta NB/\Delta R},
\end{equation}
where $F_{\rm Line}$ is the line flux in units of $\rm erg\, s^{-1}\, cm^{-2}$, $f_{\rm NB}$ and $f_{\rm R}$ are the flux densities in units of $\rm erg\, s^{-1}\, cm^{-2}\, \AA^{-1}$, and $\Delta NB$, $\Delta R$ are the FWHMs of the H$\alpha$ and R band response curves \citep{2011ApJ...726..109L, 2014ApJ...784..152A, 2018ApJ...864..145H}. 

The derived $F_{\rm Line}$ includes both H$\alpha$ and [NII] fluxes. Without metallicity or dust extinction information for this H$\alpha$ bright region, we did not apply corrections for [NII] or dust extinction and treated the measured line flux as the intrinsic flux. The H$\alpha$ flux is $1.2 \pm 0.1 \times 10^{-15}\rm \, erg\,s^{-1}\, cm^{-2}$, and the corresponding H$\alpha$ luminosity at a distance of 7.6 Mpc is $8.0 \pm 0.6 \times 10^{36} \rm \, erg\,s^{-1}$.

Using the relation $\log ({\rm SFR}_{\rm H\alpha} [{ M_\odot \rm yr^{-1}}]) = \log (L_{\rm H\alpha} \rm [erg\,s^{-1}]) -41.27$ \citep{2011ApJ...741..124H, 2012ARA&A..50..531K, 2014ApJ...784..152A}, we derived a current high-mass star formation rate
of this emitter is 4.3$\pm 0.3 \times 10^{-5}  M_\odot\, \rm yr^{-1}$. We note that this calibration assumes the H$\alpha$ emission originates from HII regions, whereas contributions from supernova remnants or planetary nebulae may introduce systematic uncertainties.

This target was also detected by GALEX, as shown in Figure \ref{stamp}. We crossmatched the catalog from the tile $\rm PS\_NGC4258\_MOS23$ and identified the source with OBJID = 2606460620868954093 (IAU Name: GALEX J121933.0+472304). 
The FUV and NUV magnitudes 
\footnote{\url{https://galex.stsci.edu/GR6/?page=explore&objid=2606460620868954093}}  are $m_{\rm FUV} = 21.60\pm 0.05$ mag, and $m_{\rm NUV} = 21.56\pm 0.05$ mag. The color $\rm mag_{\rm FUV} - mag_{\rm NUV} = 0.04$ mag implies negligible dust extinction ($A_{\rm FUV} \simeq 0$ mag). We estimated the SFR from the FUV flux using the relation ${\rm SFR}_{\rm FUV}\,[M_\odot\, {\rm yr}^{-1}] = 1.08 \times 10^{-28} L_{\rm FUV}\,[\rm erg\, s^{-1}\, Hz^{-1}]$ \citep{2007ApJS..173..267S}. At a distance of 7.6 Mpc, the derived SFR is SFR$_{\rm FUV} = 6.18\pm 0.30 \times 10^{-5} M_\odot\,\rm yr^{-1}$. The consistency between SFR$_{\rm FUV}$ and SFR$_{\rm H\alpha}$ indicates the H$\alpha$ emission is more correlated with HII region, rather than a supernova remnant.

To estimate the star formation surface density, we calculated the half-light radius of the H$\alpha$ emitter by $\rm Re_{\rm Ha} = \sqrt{Re_{\rm target}^2 - Re_{\rm PSF}^2}$, where $\rm Re_{\rm PSF} = 1.92''$. The measured half-light radius of the H$\alpha$ emitter $\rm Re_{\rm target} = 3.4''$ from SExtractor photometry catalog. So the intrinsic half-light radius is $2.8''$, or 0.10 kpc at the distance of NGC 4258. 
Then we estimate the SFR surface density as SFR/(2$\pi \rm Re_{\rm Ha}^2)$, which are $\Sigma_{\rm SFR}^{\rm H\alpha} = 6.4\times 10^{-4} M_\odot\, \rm yr^{-1} \, kpc^{-2}$ and $\Sigma_{\rm SFR}^{\rm FUV} = 9.2\times 10^{-4} M_\odot\, \rm yr^{-1} \, kpc^{-2}$.
These values are consistent with the typical star formation surface density of molecular gas regions in the Milky Way and other spiral galaxies \citep{2012ARA&A..50..531K, 2019ApJ...872...16D, 2021ApJ...908...61K}, and higher than the star forming region in the Leo HI ring \citep{2021A&A...651A..77C}.

\begin{figure}
    \centering
    \includegraphics[width=0.99\linewidth]{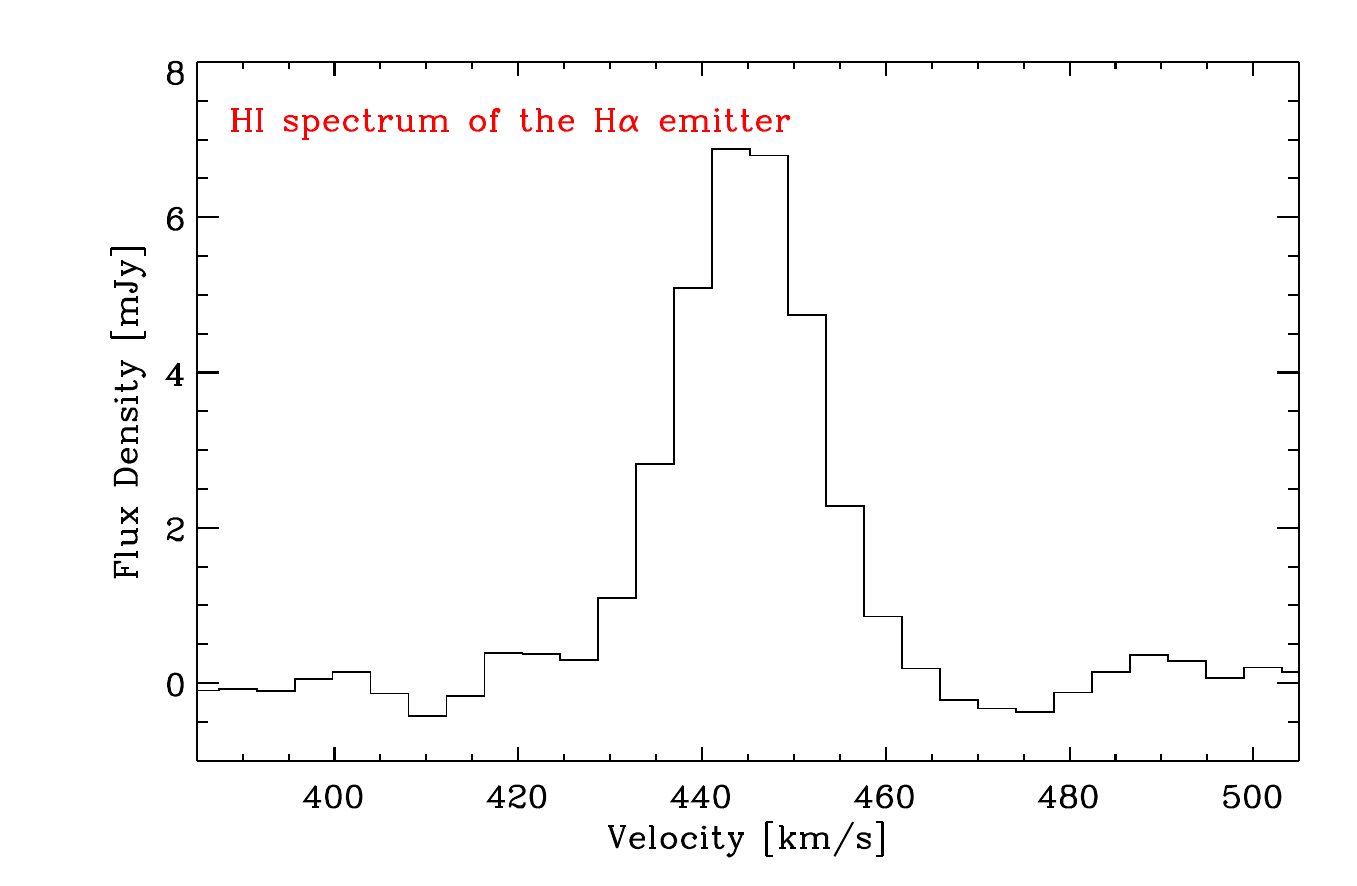}
    \caption{HI spectrum of the H$\alpha$ emitter region. The HI integrated flux is 0.12$\pm$0.01 Jy km/s, corresponding to a HI mass of $1.7 \pm 0.2 \times 10^6 M_\odot$.
    }
    \label{HIspec}
\end{figure}

The HI spectrum extracted from the cloud region within the 10$\sigma$ contour of the HI moment-0 map is shown in Figure \ref{HIspec}. The integrated flux is 0.12$\pm$ 0.01 Jy km$\rm\,s^{-1}$, and the HI mass is $1.7 \pm 0.2 \times 10^6 M_\odot$. The flux in the HI knot is barely resolved in $18.6''\times13.5''$ beam, leading to a gas surface density roughly $\Sigma_{\rm gas} \simeq M_{\rm HI}/{\rm Beam Area} = 1.6 M_\odot\,\rm pc^{-2}$. The star formation surface density and the gas surface density are consistent with Kennicutt-Schmidt relation results for low surface brightness galaxies \citep{2018ApJS..235...18L}.

\section{Discussion}

\subsection{The importance of H$\alpha$ data in this work}

UV bright targets in diffuse HI emission regions have been found in several local galaxies, such as NGC5291 \citep{2007Sci...316.1166B}, Leo ring \citep{2009Natur.457..990T}, NGC 4262 \citep{2010A&A...519A..72B}, NGC 7194 \citep{2024ApJ...961L..11P}, etc. Although the extreme blue color in UV or optical bands indicates a low metallicity, which might hint at `primordial' gas, follow up spectroscopic observations have shown a higher metallicity, indicating that the HI gas was enriched\citep{2014A&A...569A..97U, 2014ApJ...790...64R, 2021A&A...651A..77C}. The H$\alpha$ emission line observed in the spectra confirms the connection between UV and HI emission and provides constraints on the timescale of recent star formation \citep[e.g., ][]{2021A&A...651A..77C, 2024ApJ...961L..11P}.

In the case of NGC 4258, the UV emission of the H$\alpha$ emitter has been detected by GALEX and resolved by HST. The H$\alpha$ narrow band image helps constrain the redshift and provide insights into the physical properties through multi-wavelength coverage. The SFR from UV and H$\alpha$ are complementary to understanding the recent star formation in lower density outskirts of galaxies, where the stochastic nature of star formation may become more evident. Our results also highlight the potential to study star formation in galactic halos using UV and H$\alpha$ emitters with faint or no optical counterparts. 

\begin{figure}
    \centering
    \includegraphics[width=0.93\linewidth]{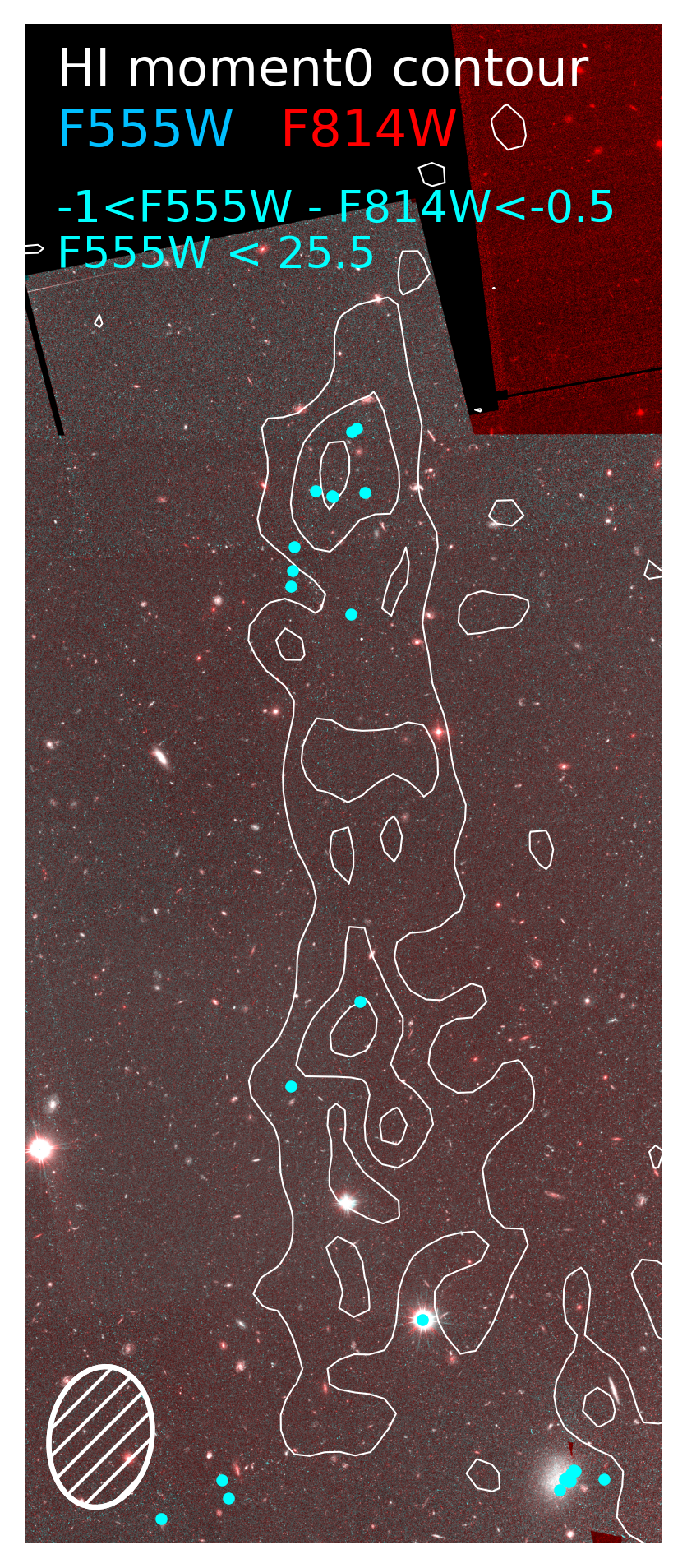}
    \caption{HST color image with wider field of view. The white contours are the HI moment0 flux. The cyan dots are the targets with $\rm -1 < F555W - F814W < -0.5$ and $\rm F555W < 25.5$, which align with the HI distribution.}
    \label{HIm0}
\end{figure}

To further validate the presence of young stellar populations in the galaxy halo, we used \textsc{SExtractor} in dual mode to perform photometry. The F555W image was used as the target detection band, while the F814W image was used for photometric measurements. We adopt default \textsc{SExtractor} parameters, except `Threshold = 5', `Seeing = 0.1'. We adopt the AUTO magnitude as the total magnitude, and the 0.06" aperture photometry to derive the F555W - F814W color. We select the blue targets with $-1 < \rm F555W - F814W < -0.5$ and $\rm F555W < 25.5$. We apply selection criteria of b/a $>$ 0.7 and FWHM $ < 0.4''$ to exclude blended targets. The selected blue targets have a median FWHM of 0.15$''$, and are marked as cyan dots in Figure~\ref{HIm0}. These sources exhibit good spatial alignment with the H I contours.
The blue targets have a median FWHM of 0.15$''$, and are highlighted as cyan dots in Figure \ref{HIm0}, which align well with the HI contour. Stellar models corresponding to $-1 < \rm F555W - F814W < -0.5$ indicate $T_{\rm eff} > 15,000$K, suggesting that the emission from these objects is dominated by massive young stars (Figure \ref{Teff}). The overlap of these extreme blue targets and the HI tail suggests recent star formation activity in the HI tail, while the lack of H$\alpha$ indicates fewer O B stars, and thus older stellar population age than the H$\alpha$ emitter \citep[e.g. Figure 9 in ][]{2021A&A...651A..77C}. On the other hand, the star formation properties such as IMF and star formation efficiency along the HI tail might be difference from HII regions in spiral disks. And mid-IR and sub-mm observations are still required to uncover the obscured star formation activity near the HI tail.

\subsection{Timescale comparison}

The F555W-F814W color of -0.8 corresponds to OB stars with an effective temperature of approximately 15,000 K, suggesting a lifetime of about 10 Myr. This result aligns well with the FUV-NUV color predictions from the Starburst99 model \citep[FUV – NUV = 0 corresponding to roughly 25 to 50 Myr, e.g., ][]{1999ApJS..123....3L, 2012MNRAS.421.3612T, 2014A&A...569A..97U}. Additionally, the H$\alpha$ emission confirms that  massive star formation has occurred within the past 10 Myr.

On the other hand, previous simulations suggest that the extended HI tail may have formed due to a close passage of a nearby galaxy \citep{2008ApJ...673..787D, 2013MNRAS.428..370S}. Assuming a rotation velocity of 200 km/s, the time required to build the HI tail, extending approximately 10 kpc, would be around 50 Myr. This timeline is consistent with the possibility that star formation occurs concurrently with the creation of the HI tail. The HI tail around NGC 4258 is not primordial in composition, but pre-enriched within the galactic disk before being stripped by tidal interactions. For the majority of this process, star formation activities—such as the formation of giant molecular clouds, gas cooling, and subsequent star formation—occur outside the main galaxy.

\subsection{The Fate of the HI Tail: Potential Dwarf Galaxy Formation and CGM Enrichment}

The emergence of the HI tail may result from weak interactions with the satellite galaxy NGC 4258 \citep{2008ApJ...673..787D} or faint companion galaxies revealed by recent FAST observations \citep{2021ApJ...922L..21Z}. If the HI tail continues to grow and several star clusters remain gravitationally bound, this gas structure could evolve into a dwarf galaxy with little dark matter \citep{2013MNRAS.429.1858D}. Such a system would likely be diffuse due to its weak gravitational potential. Subsequently, dynamical processes like three-body interactions could eject stars from the system over time, progressively lowering its stellar mass and potentially driving its evolution into one of the faintest known dwarf galaxies, similar to Ursa Major III or UNIONS 1 \citep{2019ARA&A..57..375S, 2024ApJ...965...20E}. Star-forming regions within the HI tail, resembling those in tidal dwarf galaxies \citep{2009AJ....137.4561B, 2010AJ....140.2124B, 2012ASSP...28..305D, 2015A&A...584A.113L} but with lower total stellar mass, support this scenario.

Furthermore, the ongoing star formation within the HI tail may also enrich the circumgalactic medium \citep[e.g.,][]{2011Sci...334..948T, 2017ARA&A..55..389T, 2023Sci...380..494Z}. The young blue stars in the H$\alpha$ region are likely embedded in a low-density interstellar medium, allowing the resulting supernovae to remain in the free-expansion phase for a longer time. If the ejecta can maintain a velocity of $\sim$10,000 km s$^{-1}$, the blast wave could transport metal-enriched ejecta to distances of $\sim$100 kpc within $\sim$10 Myr, directly enriching the circumgalactic medium.

Neutral hydrogen in the outskirts of galaxies is particularly vulnerable to disruption during interactions with neighboring systems. Such interactions can both redistribute and compress gas, triggering star
formation that results in young stars far out in the galaxy's halo or
extended stellar disk, thereby enhancing the richness and structural complexity of the galaxy.

\section{Summary} \label{sec:cite}

In this study, we report the detection of an H$\alpha$ emitting region in the eastern HI tail of NGC 4258. Combining multi-wavelength data including UV, H$\alpha$, and optical observations, we investigate basic properties of the H$\alpha$ region. We obtain the star formation rate derived from H$\alpha$ as $4.3 \pm 0.3 \times 10^{-5} M_\odot \,\rm yr^{-1}$, which is consistent with the SFR estimated from FUV emission ($6.2\pm 0.3 \times 10^{-5} M_\odot \,\rm yr^{-1}$). High resolution optical images from HST reveal several blue targets within the H$\alpha$ emission region, which are about 25.5 AB mag in F555W, corresponding to about $\sim 10^3L_\odot$, and are likely to be  
a young open cluster candidate, while more observations are still needed to identify the nature of the H$\alpha$ emitter. The extreme UV and optical color (FUV-NUV $\simeq$ 0; F555W - F814W $\simeq$ -0.6) and the H$\alpha$ emission constrain that the star formation activity occurred within 10 Myr, which is comparable to the formation timescale of the HI tail. So the star formation is likely to have started during the emergence of the HI tail. The recent star formation activity further demonstrates that HI tail around galaxies would increase the stellar halo mass, and the metallicity in the circumgalactic medium. 

 Although only a single massive star forming region was identified
within the HI tails of NGC 4258 in this study, our results highlight the importance of deep narrow band images for identifying the very few regions of recent star formation in the galaxy halo. Follow-up observations such as optical spectroscopy,  mid-IR, CO \citep[e.g., ][]{2023A&A...671A.104C} and high-resolution HI observations are crucial to better understand the nature of the emitter and its role in galaxy formation and evolution.

\begin{acknowledgments}

We thank the referee for the careful reading and detailed comments. The step-by-step suggestions have greatly helped us not only improve the quality of the manuscript, but also deepen our understanding of the subject. We thank the NOWT observation assistants and Hubiao Niu, Chunhai Bai, Shuguo Ma at the Nanshan Station of Xinjiang Astronomical Observatory for their assistance during the observations. We also thank Zhenya Zheng, Jie Wang and Jing Wang for their valuable suggestions and insightful discussions during the development of this project.

This work is supported by Tian-shan Talent Training Program under No. 2023TSYCLJ0053. This work is supported by the National Natural Science Foundation of China, No. 11803044, 11933003, 12173045. This work is sponsored (in part) by the Chinese Academy of Sciences (CAS), through a grant to the CAS South America Center for Astronomy (CASSACA). C.C. is supported by Chinese Academy of Sciences South America Center for Astronomy (CASSACA) Key Research Project E52H540301. PW gratefully acknowledges the support from Chinese Academy of Sciences (CAS) ``Light of West China'' Program (No. 2021-XBQNXZ-029). This work is supported by the China Manned Space Program with grant no. CMS-CSST-2025-A07.

Some of the data presented in this paper were obtained from the Mikulski Archive for Space Telescopes (MAST) at the Space Telescope Science Institute. The specific observations analyzed can be accessed via \dataset[https://doi.org/10.17909/jhga-gy40]{https://doi.org/10.17909/jhga-gy40}. STScI is operated by the Association of Universities for Research in Astronomy, Inc., under NASA contract NAS5–26555. Support to MAST for these data is provided by the NASA Office of Space Science via grant NAG5–7584 and by other grants and contracts.

\end{acknowledgments}

\begin{contribution}

The project was conceived by Cheng Cheng and Jia-Sheng Huang. Cheng Cheng performed the data analysis and prepared the manuscript. Zhixiang Zhang constructed the stellar models to estimate stellar parameters. Guojie Feng, Letian Wang, Abdusamatjan Iskandar, Shahidin Yaqup assisted with the observations. All authors contributed to the discussion and interpretation of the results.


\end{contribution}

%

\facilities{NOWT:1m, HST(ACS), GALEX, WRST}


\software{astropy \citep{2013A&A...558A..33A, 2022ApJ...935..167A, 2018AJ....156..123A},  
          Source Extractor \citep{1996A&AS..117..393B},
          La COSMIC \citep{2012ascl.soft07005V},
          Astrometry.net \citep{2010AJ....139.1782L},
          SWARP \citep{2010ascl.soft10068B}
          }


\appendix

\section{Stellar effective temperature and F555W-F814W color}

We estimate the relation between the blue color and the effective temperature with `BT-COND' stellar spectra model \footnote{\url{http://svo2.cab.inta-csic.es/theory/newov2/index.php?models=bt-cond}} \citep{2011ASPC..448...91A, 2012RSPTA.370.2765A, 2006MNRAS.368.1087B, 2011SoPh..268..255C}. We set the $\log g = 2.5$ and 3, and show the NUV-R, F555W-F814W color as a function of $T_{\rm eff}$ in Figure \ref{Teff}. The F555W - F814W color lower than -0.5 need extremely hot and massive stars. The non-detection in DECaLS R band image also indicates a NUV - R color lower than -2,
providing additional support for our conclusion that the continuum
emission originates from a young, massive stellar population.

While the H$\alpha$ emitter is not resolved in the GALEX or DECaLS R-band images, we analyze the color distribution of stars with $T_{\rm eff} > 15{,}000$ K in Figure~\ref{starcolor}, using FUV - NUV versus NUV - R and NUV - F555W colors. We find that the observed NUV - R and NUV - F555W colors are still bluer than the stellar population model predictions when FUV - NUV is close to zero. Since both the GALEX and R-band images lack sufficient spatial resolution to isolate the H$\alpha$-emitting region, high-resolution HST imaging in bluer bands will be essential to better constrain the underlying stellar population and the possible dust extinction.

\setcounter{figure}{0}
\renewcommand{\thefigure}{A\arabic{figure}}

\begin{figure*}
    \centering
    \includegraphics[width=0.95\linewidth]{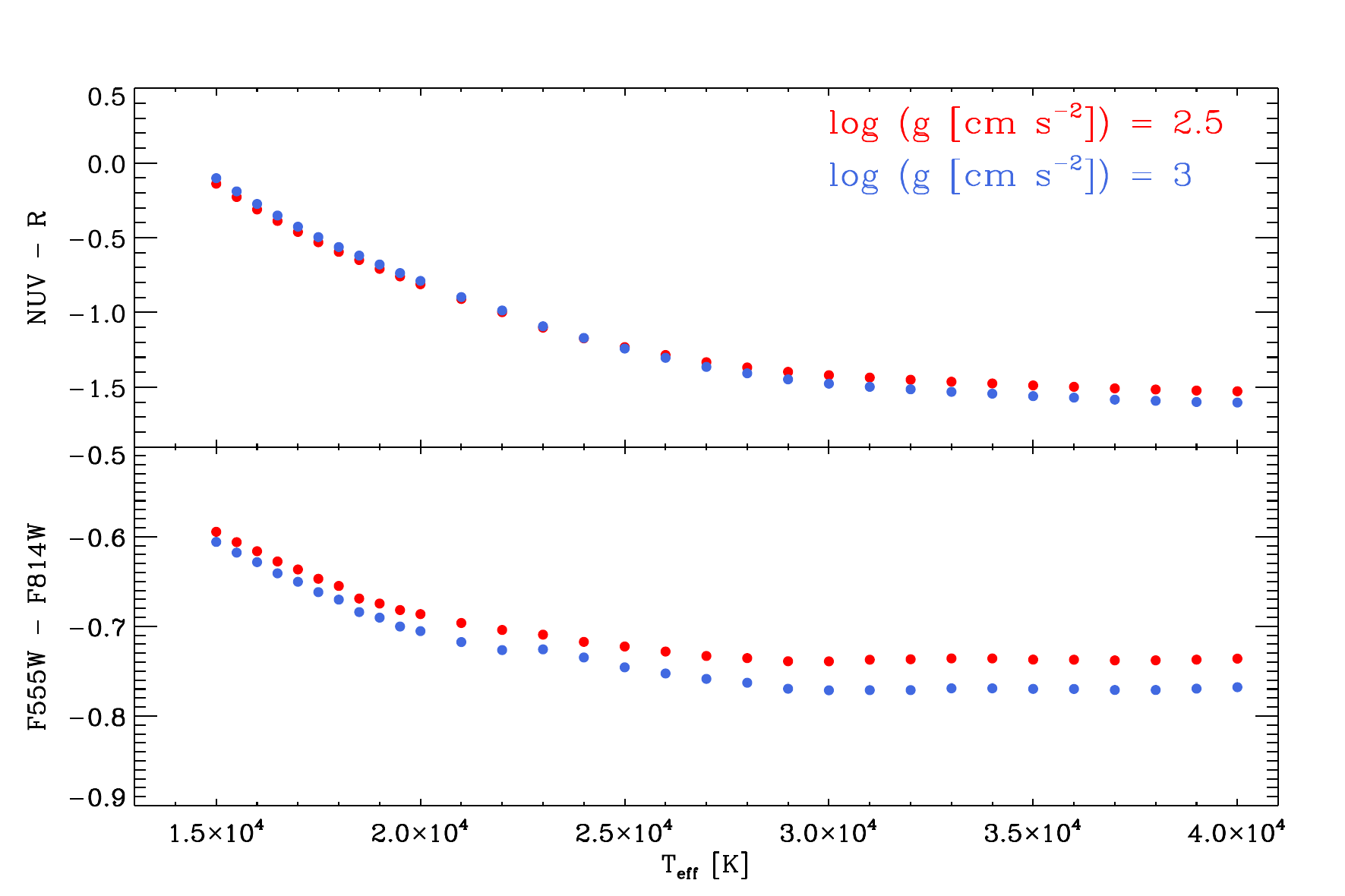}
    \caption{$T_{\rm eff}$ and NUV - R color (top panel), F555W-F814W color (bottom panel) for stars with $\log \rm g\, [cm\,s^{-2}] = 2.5$ and 3. So the typical $T_{\rm eff}$ for a star with F555W-F814W lower than -0.5 should be above 15{,}000 K. 
    The H$\alpha$ emitter is not detected in the DECaLS R-band image, with an R-band magnitude fainter than 23 (depending on the photometric aperture). Then the upper limit on the NUV $-$ R color is lower than $-2$, indicating the presence of a young, massive stellar population.}
    \label{Teff}
\end{figure*}

\begin{figure*}
    \centering
    \includegraphics[width=0.9\linewidth]{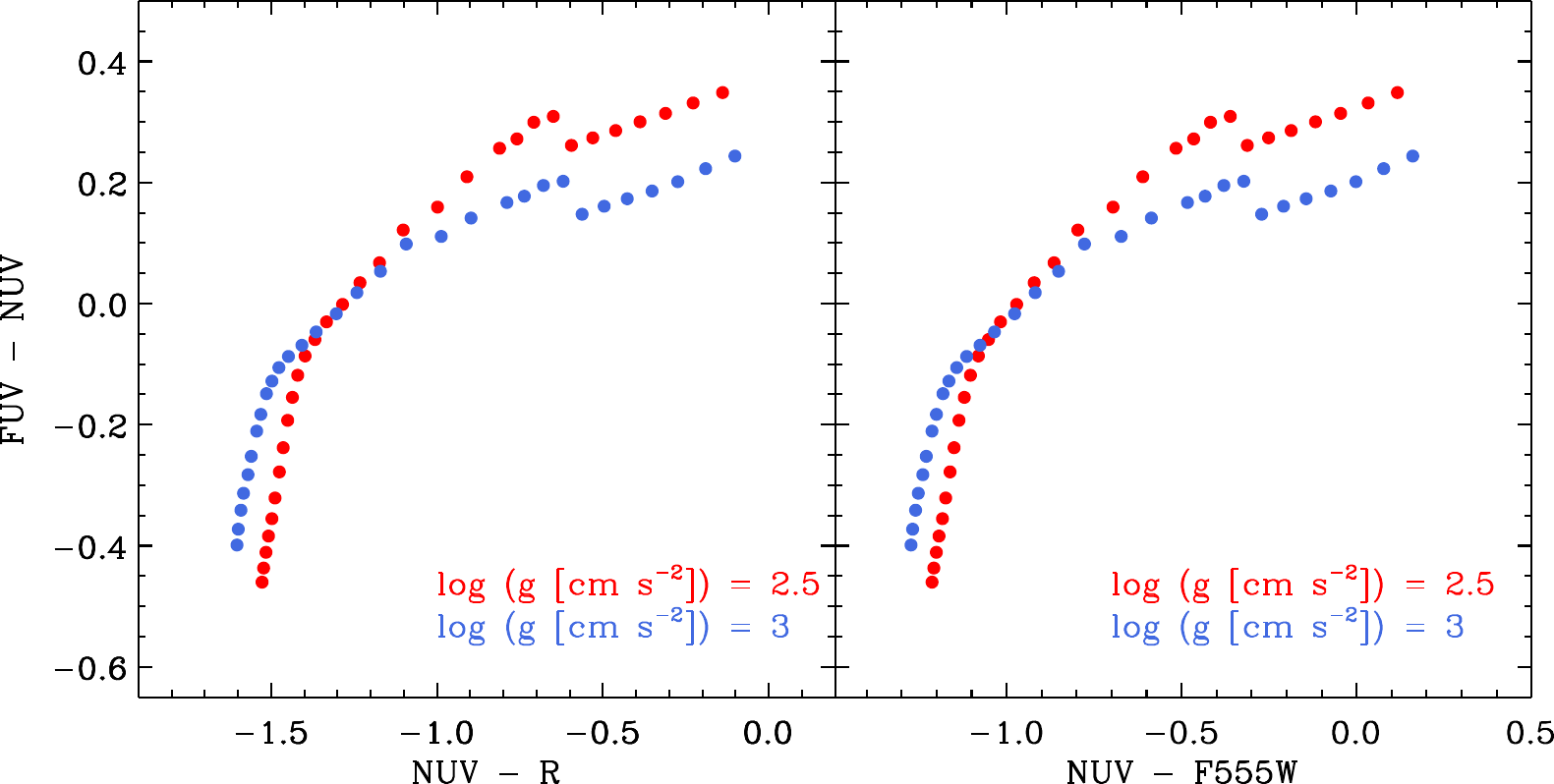}
    \caption{FUV-NUV vs NUV-R (left panel) and NUV - F555W (right panel) color for stars with $\log \rm g\, [cm\,s^{-2}] = 2.5$ and 3, and $T_{\rm eff} > 15{,}000$ as in Figure \ref{Teff}.
    }
    \label{starcolor}
\end{figure*}





\end{document}